\documentclass[usenatbib]{basi}
\usepackage[T1]{fontenc}
\usepackage[british]{babel}
\usepackage[varg]{txfonts}

\usepackage{rotating}
\usepackage{dcolumn}

\newcommand{\simnot}{\mathord{\sim}}

\begin{document}
\title[The spectral structure and energetics  of powerful radio sources]{The spectral structure and energetics of powerful radio sources}
\author[J.~J.~Harwood et~al.]%
       {J.~J.~Harwood$^1$\thanks{email: \texttt{jeremy.harwood@physics.org}},
       M.~J.~Hardcastle$^{1}$, J.~H.~Croston$^2$, A. ~Stroe$^3$, R. ~Morganti$^{4,5}$ and E. ~Orr{\`u}$^4$\\
       $^1$School of Physics, Astronomy and Mathematics, University of Hertfordshire, College Lane, Hatfield, Hertfordshire AL10 9AB, UK\\
       $^2$School of Physics and Astronomy, University of Southampton, Southampton SO17 1BJ, UK\\
       $^3$Leiden Observatory, Leiden University, PO Box 9513, 2300 RA Leiden, The Netherlands\\
       $^4$Netherlands Institute for Radio Astronomy (ASTRON), Postbus 2, 7990 AA Dwingeloo, The Netherlands\\
       $^5$Kapteyn Astronomical Institute, University of Groningen, Post Office Box 800, 9700 AV Groningen, The Netherlands}

\pubyear{2014}
\volume{00}
\pagerange{\pageref{firstpage}--\pageref{lastpage}}
%\status{submitted}

\date{Received --- ; accepted ---}

\maketitle
%------------------------------------------------------------------------------%
% abstract and keywords                                                        %
%------------------------------------------------------------------------------%
\label{firstpage}

\begin{abstract}
Determining the energy spectrum of an electron population can give key insights into the underlying physics of a radio source; however, the lack of high resolution, broad-bandwidth observations has left many ambiguities in our understanding of radio galaxies. The improved capabilities of telescopes such as the JVLA and LOFAR mean that within the bandwidth of any given observation, a detailed spectral shape can now be produced. We present recent investigations of powerful FR-II radio galaxies at GHz and MHz frequencies and show for the first time their small-scale spectral structure. We highlight problems in traditional methods of analysis and demonstrate how these issues can now be addressed. We present the latest results from low frequency studies which suggest a potential increase in the total energy content of radio galaxy lobes with possible implications for the energetics of the population as a whole.

\end{abstract}


%------------------------------------------------------------------------------%
% main text of the paper, using \section, \subsection, \subsubsection          %
%------------------------------------------------------------------------------%
\section{Introduction}
\label{introduction}

In principle, a region emitting synchrotron radiation will preferentially cool higher energy electrons leading to a steeper, more strongly curved spectrum in older regions of plasma. Models of this `spectral ageing' have become a commonly used tool when describing the processes involved in the lobes of FR-I and FR-II type radio galaxies. Historically, investigations of spectral ageing in radio sources have largely been limited by the fact that only a few narrow-band frequencies have been available. However, the new generation of radio telescopes has changed this situation dramatically. The increased sensitivity, bandwidth and dynamic range provided by instruments such as LOFAR and the JVLA now allow for spectral curvature to be determined across an entire frequency band of a single pointing. This type of detailed spectral analysis is set to become standard practice, hence it is vital that methods are developed to handle new broadband radio observations. The `Broadband Radio Analysis ToolS' (BRATS) was therefore developed to provide a range of analysis, statistical and model fitting tools for broadband radio data \citep{harwood13}. Within these proceedings, we detail how these new methods have been applied to observations at GHz frequencies, and describe our current research at low-frequencies as part of the LOFAR nearby AGN key science project (KSP).

\section{Spectral ageing at GHz frequencies}
\label{ghzfreq}

\begin{figure}
\centerline{\includegraphics[width=6cm]{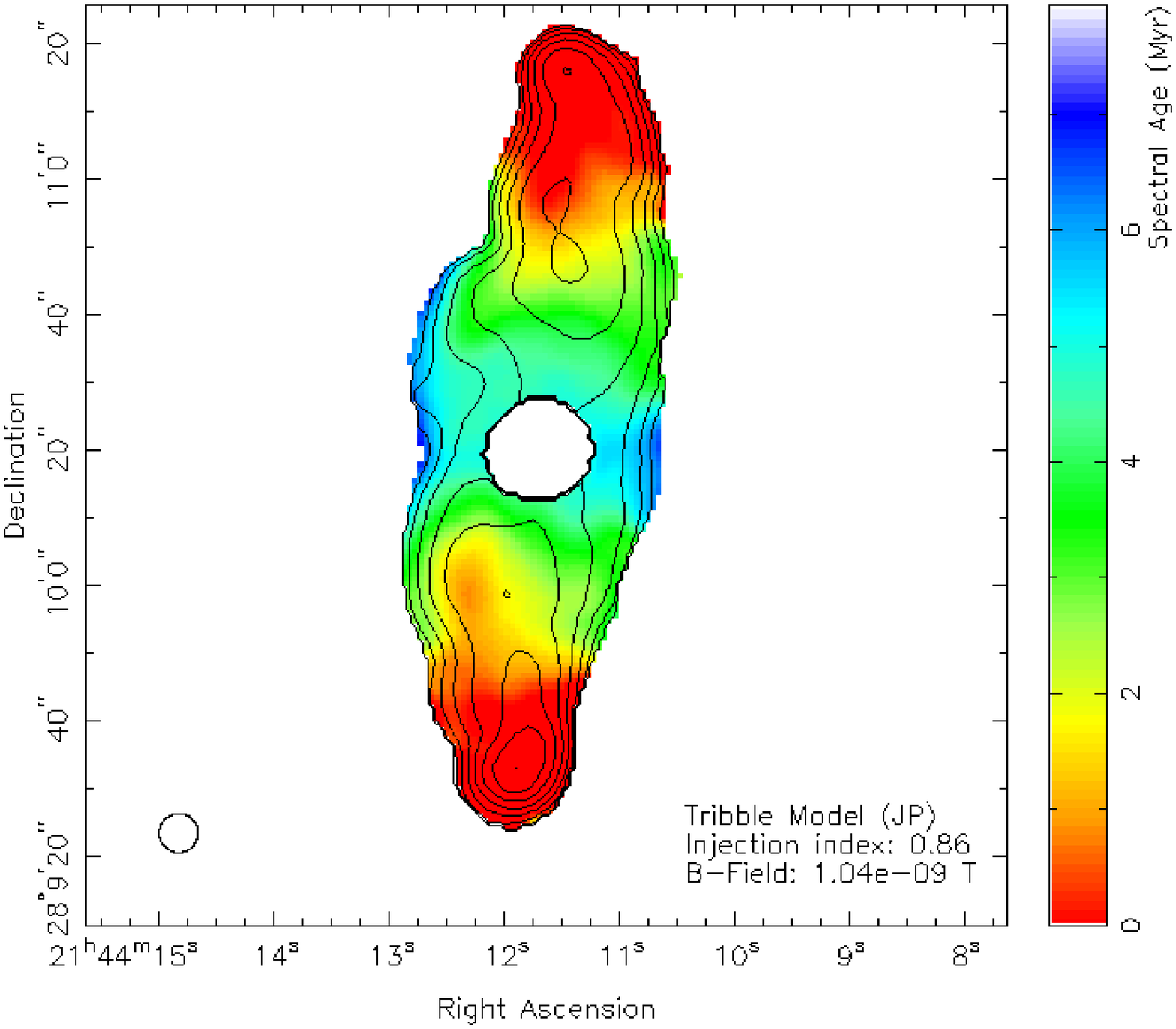} \qquad
            \includegraphics[width=6cm]{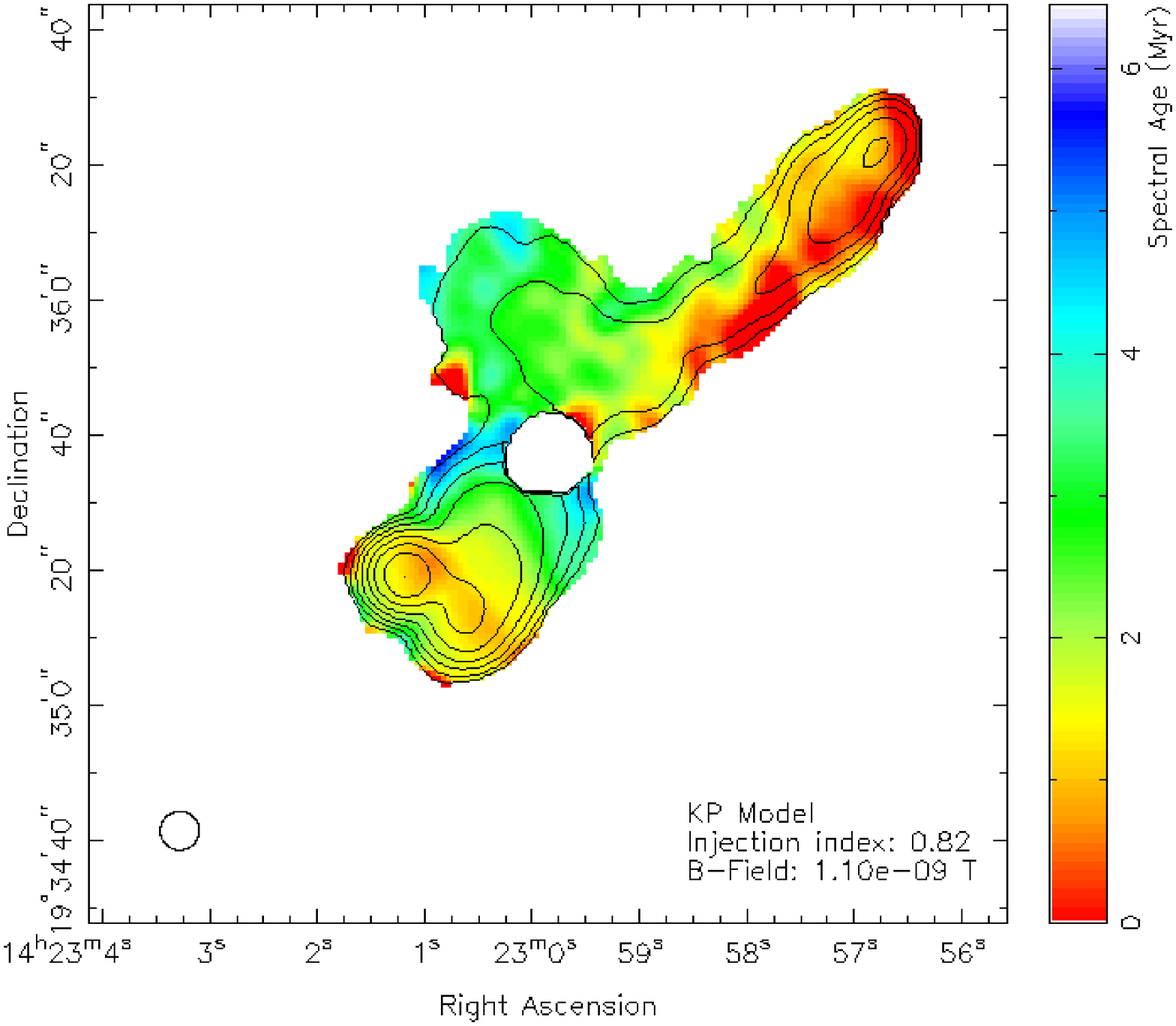}}
\caption{Spectral age maps of 3C436 (left) and 3C300 (right) fitted using the Tribble model with $\alpha_\mathrm{{inj, 3C436}} = 0.86$, $\alpha_\mathrm{{inj, 3C300}} = 0.82$ respectively.\label{specagemaps}}
\end{figure}

The two most commonly used models of spectral ageing are those first proposed by \citet{kardashev62} and \citet{pacholczyk70} (the KP model) and \citet{jaffe73} (the JP model). Fitting of these models has traditionally relied on the use of large spatial regions in order to obtain the required signal to noise levels; however, as the time since initial particle acceleration is known to vary as a function of position within the lobes of radio galaxies, this leads to the analysis of a superposition of spectral ages. This complicating factor has previously assumed to be negligible, but as the spectral structure of the extended emission in powerful radio sources is largely unknown on small spatial scales, its validity is not immediately evident. By applying the methods developed by \citet{harwood13} to recent JVLA observations, we show that it is now it is now possible to consider these sources on much smaller spatial scales. Observations were made of 2 bright, nearby FR-II sources selected from the 3CRR sample \citep{laing83}; 3C436 and 3C300. Using the BRATS software, keys model parameters were determined and spectral ageing models fitted on a pixel by pixel basis. The importance of considering spectra on these small scales is immediately evident from the maps of spectral age shown in Figure \ref{specagemaps}. We see clear cross-lobe variations, particularly in the case of 3C300, where the low-age regions stretch along the length of the jet are likely due to previously unknown jet interaction. The validity of negligible age variation when using classical methods of analysis over large regions are therefore unlikely to be valid in a significant number of cases.

Both the KP and JP models of spectral ageing assume a fixed magnetic field strength and a single injection electron energy distribution. This so-called injection index is commonly assumed to take values between $\simnot\,0.5$ and $0.7$; however, from our analysis using direct fitting of these models, both sources take values in excess of $0.8$. One plausible explanation for these high injection indices is a lack of curvature in the models at low-frequencies. Testing of a modified JP model (the Tribble model) \citep{tribble93, hardcastle13, harwood13} which instead assumes a Gaussian random magnetic field causing additional low-frequency curvature, we find no difference in the derived value. Current research into spectral ageing in diffuse, galaxy cluster radio relics also hints at a similar deviation from expected values. It is therefore unlikely that this deviation is model dependent. Recent papers by \citet{weeren10} and \citet{stroe13} into the radio relic CIZA J2242.8+5301 measure an injection index of $\simnot\,0.6$; however, our current works (Stroe et al., in prep) find the injection index to be $\simnot\,0.77$. Given that the dynamics of radio relics are much simpler than those of FR-II galaxies, and are also thought to have well ordered magnetic fields, this is a somewhat surprising result.

A second key issue which also arises from our analysis of FR-II galaxies is that of the reliability of observed spectral ages as an estimate of the true source age. It has long been known that the spectral ages of powerful radio sources are underestimated when compared to those derived dynamically (e.g. \citealp{eilek96}). Confirmation of this disparity using a method which is independent of previous findings supports a picture in which this is not simply an artefact of the methods employed, but a critical problem which must be addressed. We consider a range of options in the context of spectral ageing such as non-equipartition magnetic fields, but find that no one parameter can fully account for the observed difference in age. Results from investigations of FR-I galaxies \citep{heesen14}, along with recent MHD simulations of FR-II galaxies by Hardcastle (private communication), show that mixing of electron populations may play a key role in determining the observed spectra, hence age, of radio galaxies. It is therefore likely that new or modified models of spectral ageing will be required to resolve this disparity. Regardless of these outstanding issues we find that spectral ageing in radio galaxies does appear to have a basis in physical reality. We suggest that new models of spectral ageing, such as the Tribble model, potentially provide a more accurate and physically realistic description of these complex sources.

\section{Spectral structure and energetics at low-frequencies}\label{lowfreq}

On resolved scales, the diffuse emission from the lobes of FR-II radio galaxies remains largely unexplored at low-frequencies. As part of the LOFAR nearby AGN key science project, observations were therefore made of the nearby FR-II source 3C223 between 30 and 240 MHz. Although at only the early stages of analysis, preliminary results are promising. One of the primary limiting factors with such spectral analysis is determining an accurate flux scale to constrain any spectral curvature present within the lobes of FR-II galaxies. When compared to integrated fluxes extrapolated from 74 MHz using the flux density and spectral index of \citet{orru10}, the recovered values are accurate to within 5 per cent with an underestimation of flux found at LBA and an overestimation at HBA frequencies. This is consistent with the expectation that spectral curvature is present at these wavelengths. Further analysis using the methods employed in Section \ref{ghzfreq} is required to confirm if such curvature is truly present, but these results give the first insight into the detailed spectrum at these low-frequencies.

A preliminary analysis has also been undertaken into the energetics of 3C223. X-ray inverse-Compton measurements have previously been used to constrain the total energy content of the lobes of radio galaxies; however, it has been necessary to make assumptions about the spectral profile at radio wavelengths. With these new observations, we are now able to revisit these results and better constrain the models at low-frequencies. Using the methods of \citet{croston04}, the integrated flux values for the northern and southern lobes of 3C223 were added to extend the frequency coverage of the original values. Fitting these revised measurements to the inverse-Compton models, we find a significant increase in the total energy content, which in the case of the northern lobe may be up to twice the originally derived value. This increase is due to a steeper spectrum at low-frequencies than had been originally assumed, leading to more energy being stored in the low-energy electron population. Interestingly, this assumption has also been applied to later studies \citep{croston05} over a much larger sample, hence, if these steep spectra are found to be common in all powerful radio galaxies, the energy of the population as a whole will be significantly increased. Further analysis is required to confirm these preliminary findings, but hint at a potential shift in our understanding of the energetics of radio galaxies.

%------------------------------------------------------------------------------%
% bibliography: produced from ADS using custom format of                       %
%                                                                              %
%     %z132 \\bibitem[%\2%(y)%\3m]%{R}\n   %\8.1g,%\Y,%\q,%\V,%\ p             %
%------------------------------------------------------------------------------%

\label{lastpage}
%------------------------------------------------------------------------------%
\end{document}